\documentclass[fleqn,usenatbib]{mnras}

\usepackage{newtxtext,newtxmath,breqn}
\usepackage{xcolor}

\usepackage[T1]{fontenc}

\DeclareRobustCommand{\VAN}[3]{#2}
\let\VANthebibliography\thebibliography
\def\thebibliography{\DeclareRobustCommand{\VAN}[3]{##3}\VANthebibliography}
\DeclareRobustCommand{\VANDER}[3]{#2}
\let\VANDERthebibliography\thebibliography
\def\thebibliography{\DeclareRobustCommand{\VANDER}[3]{##3}\VANDERthebibliography}
\DeclareRobustCommand{\DE}[3]{#2}
\let\DEthebibliography\thebibliography
\def\thebibliography{\DeclareRobustCommand{\DE}[3]{##3}\DEthebibliography}

\usepackage{graphicx}	
\usepackage{amsmath}	
\usepackage{color}

\defcitealias{Zibetti:2017aa}{Z17}
\defcitealias{BC03}{BC03}

\newcommand{\dagen}{$\Delta {\rm age_{n}}$}
\newcommand{\dagenmin}{$\Delta {\rm age_{n,min}}$}
\newcommand{\invAA}{\AA$^{-1}$}

\title[Age resolution of stellar populations]{On the maximum age resolution achievable through stellar population synthesis models}

\author[S. Zibetti et al.]{
Stefano Zibetti,$^{1,2}$\thanks{E-mail: stefano.zibetti@inaf.it (SZ)}
Edoardo Rossi,$^{1,2}$
and Anna R. Gallazzi$^{1}$
\\
$^{1}$INAF-Osservatorio Astrofisico di Arcetri, Largo Enrico Fermi 5, I-50125 Firenze, Italy\\
$^{2}$Universit\`a degli Studi di Firenze, Dipartimento di Fisica e Astronomia, Firenze, Italy
}

\date{Accepted 2024 January 08. Received 2023 December 07; in original form 2023 July 19}

\pubyear{2024}

\begin{document}
\label{firstpage}
\pagerange{\pageref{firstpage}--\pageref{lastpage}}
\maketitle

\begin{abstract}
As the reconstruction of the star-formation histories (SFH) of galaxies from spectroscopic data becomes increasingly popular, we explore the best age resolution achievable with stellar population synthesis (SPS) models, relying on different constraints: broad-band colours, absorption indices, a combination of the two, and the full spectrum. We perform idealized experiments on SPS models and show that the minimum resolvable relative duration of a star-formation episode (time difference between 10\% and 90\% of the stellar mass formed divided by the median age) is never better than 0.4, even when using spectra with signal-to-noise ratio (SNR) larger than 100 per \AA. Typically, the best relative age resolution ranges between 0.4 and 0.7 over most of the age-metallicity plane, corresponding to minimum bin sizes for SFH sampling between 0.15 and 0.25 dex. This resolution makes the spectroscopic exploration of distant galaxies mandatory in order to reconstruct the early phases of galaxies' SFHs. We show that spectroscopy with ${\rm SNR} \gtrsim 20\,$\AA$^{-1}$ is essential for good age resolution. Remarkably, using the full spectrum does not prove significantly more effective than relying on absorption indices, especially at ${\rm SNR}\lesssim 20$\invAA. We discuss the physical origins of the age resolution trends as a function of age and metallicity, and identify the presence of maxima in age resolution (i.e. minima in measurable relative time duration) at the characteristic ages that correspond to quick time variations in spectral absorption features. We connect these maxima to bumps commonly observed in reconstructed SFHs.
\end{abstract}




\section{Introduction}\label{sec:intro}

A galaxy's spectral energy distribution (SED), throughout the ultra-violet (UV), visible and near infra-red (NIR) wavelength range, is chiefly determined by the emission of multiple generation of stars, which may follow a variety of star-formation histories (SFH) and chemical-enrichment histories (also alternatively described as a 2-dimensional age-metallicity relation, AMR). Thanks to more than fifty years of contributions by a vast community, pioneered by Beatrice Tinsley (see her fundamental \citeyear{tinsley_1980} review), we can rely on very well established methods to produce finely detailed forward models of the UV-visible-NIR emission of a galaxy, whose accuracy and reliability are generally best in the visible range. These models are produced starting from an assumed SFH or AMR, by means of stellar population synthesis (SPS) models, and may now include the effects of dust and nebular emission by the ionized gas \citep[see the comprehensive review by][figure 1 in particular]{conroy_2013}. A reliable method for inverting the spectral information that is directly observable from a galaxy into its SFH (or AMR) is a holy grail of the extragalactic astrophysics, as it would allow to derive a substantial part of the evolutionary history of a galaxy just by analysing its spectrum. Unfortunately, on the one hand, a large number of systematic uncertainties still plague the forward modelling of galaxy spectra. On the other hand, possibly more worrisome, even if we neglect the systematic uncertainties in the synthesis models, it has to be realized that the spectral inversion is an essentially ill-conditioned problem, as demonstrated in detail by \cite{Ocvirk+2006}, due to the degeneracy affecting the spectra of different stellar populations.

Despite these well known limitations, methods and algorithms have been devised to tackle this problem, adopting different statistical approaches and physical assumptions. One of the key features that distinguish between spectral inversion codes is the choice of assuming parametric or non-parametric SFHs. In the former case, the SFH is assumed to be an analytic function whose shape is controlled by a (small) number of parameters that are determined either via likelihood maximization or by means of a probability distribution function (PDF), e.g. in a Bayesian framework \citep[e.g. \textsc{bagpipes}][]{carnall+2018}. As opposed, in fully non-parametric fitting codes the reconstructed SFH is represented by the weights of stars in bins of age (and metallicity). While the base models of simple stellar populations (SSP) that are combined to create the composite stellar population representing a galaxy can be computed with a very high age sampling, it is obvious that the actual age resolution that is possible to recover via inversion is very much limited. In other words, the width of independent age bins in a reconstructed SFH can not be small at will, but it is limited by the amount of information that is possible to extract from a spectrum. In some codes this problem is not considered and the binning just follows the input grid of SSPs \citep[e.g. \textsc{starlight}][]{STARLIGHT}. Other codes cope with the physical limitation of age binning by implementing solutions which range from adopting information criteria to limit the number of components \citep[e.g. \textsc{firefly}][]{Wilkinson+2017}, to regularizing penalization \citep[e.g. \textsc{pPXF}][]{Cappellari:2017}, iterative binning refinement \citep[e.g. \textsc{VESPA}][]{Tojeiro+2007}, and ad hoc devised binning based on simulated datasets \citep[e.g. \textsc{prospector}][]{Leja+2017,Leja+2019}.

There are in particular two astrophysical problems in which knowing the achievable age resolution is key in order not to over-interpret the data: the reconstruction of the ancient phases of star formation for (relatively) nearby galaxies, and the identification of short-duration bursts of star formation along the SFH of a galaxy (which can be used e.g. as signpost of accretion events or interactions).




In this paper we set to investigate the theoretical age resolution limits that can be attained by using the integrated photometric and spectroscopic properties of a composite stellar population as constraints. We define as ``age resolution'' the minimum time duration of an extended SFH that is able to produce a significant spectral change with respect to an instantaneous burst of the same age as the median age of the stars in the extended SFH. We will explore how the age resolution varies as a function of median stellar age and metallicity, for different SNR of the spectral constraints. 

We stress that this paper is not concerned with the actual performance and time-resolution of different SFH inversion codes on realistic galaxy spectra, rather we aim at setting the theoretical limits in age resolution that derive from the \emph{limited} and \emph{partly degenerate} information about stellar populations, which is encoded in galaxy spectra. For this reason, we analyze only ideal cases in which: \emph{i}) the SFH is given by a smooth analytic function, excluding any burst; \emph{ii}) the metallicity is uniform and constant throughout the SFH; \emph{iii}) there is no dust resulting in extinction or differential attenuation; \emph{iv}) the IMF is constant and universal. As we will discuss in Sec. \ref{sec:discussion} and Appendix \ref{app:dagen_recovery}, the ability to constrain the SFH duration in realistic cases is expected to be much worse than in the idealized cases considered to set the age resolution limits.

We will consider different sets of spectro-photometric constraints over the visible spectral range, namely: \emph{i}) absorption indices \citep[as in][]{gallazzi+05}, \emph{ii}) broad-band colours, \emph{iii}) a combination of absorption indices and broad-band colours \citep[as in][]{Zibetti:2017aa}; \emph{iv}) the full spectrum ($R\simeq2000$ resolution\footnote{$R\simeq2000$ is approximately the typical effective spectral resolution for the spectroscopy of external galaxies, resulting from the combination of instrumental resolution and intrinsic Doppler broadening due to velocity dispersion of the order of some 100 km s$^{-1}$.}), over two different spectral ranges ($\lambda\lambda 3800-5600$\AA~ ``restricted'' and $\lambda\lambda 3500-7000$ \AA~ ``extended''), using either the rectified spectrum (i.e. the spectrum normalized by the continuum shape) or the spectrum as is. These different sets of constraints embrace essentially all possible combinations of spectro-photometric constraints: from broad-band SED, to a limited selection of reliable/well-modeled absorption features, to the full wealth of spectral features at the moderate resolution of typical extragalaxtic surveys, possibly complemented by information on the SED shape.

Given the ideal nature of this experiment and the fact that we are not interested in absolute numbers for age estimates rather in relative differences, we consider only one suite of simple stellar population (SSP) models to build our composite stellar populations. Our analysis is based on the SSPs built with the 2016 version of the \citet[][hereafter BC03]{BC03} stellar population synthesis code, using the MILES spectral libraries \citep[][]{Sanchez-Blazquez:2007, Falcon-Barroso:2011aa} and adopting the ``Padova 1994'' evolutionary tracks and the \cite{chabrier03} initial mass function (see \citetalias{BC03} for the complete list of references about the evolutionary tracks and isochrones). These SSP models are publicly available at \url{http://www.bruzual.org/bc03/Updated_version_2016/}. The adoption of other models is not expected to change the main results of this work, as will be discussed in Sec. \ref{sec:discussion}.

This paper is organized as follows. In Sec. \ref{sec:method} we introduce the operational definitions of duration of the SFH and of age resolution and present the methodology employed to measure them over a range of different ages, metallicity and observational conditions. In Sec. \ref{sec:results_indx_phot} we present the results obtained when absorption indices and broad-band photometric fluxes (either by themselves or combined) are used as constraints, while in Sec. \ref{sec:results_full_spec} we present the results based on full spectral analysis. In Sec. \ref{sec:discussion} we will discuss the validity and limitations of our results (Sec. \ref{sub:limitations}), compare them with other studies in the literature (Sec. \ref{sub:disc_literature}), provide interpretations of the age resolution trends in terms of stellar evolution (Sec. \ref{sub:disc_physical}) and related spectral features (Sec. \ref{sub:disc_spec_feat}), and discuss the implications for SFH reconstruction (Sec. \ref{sub:disc_SFHreconstr}). A summary and the concluding remarks are given in Sec. \ref{sec:conclusions}.

\section{Method and stellar population models}\label{sec:method}
In this section, first of all we define the ``relative extension'' \dagen~of a SFH as a model-independent quantity and, by means of that, we operationally define the ``age resolution'' (Sec. \ref{sub:dagen}). We then describe the set of model SFHs and the resulting spectra and spectral diagnostics employed in our analysis (Sec. \ref{sub:library}). The basic idea of our experiment is to map the minimum ``relative extension'' \dagenmin~that results in statistically significant differences with respect to the spectra produced by SFHs of \dagen$\simeq 0$ and equal median age, as a function of median age, metallicity and SNR of the observations. In order to quantify the statistical significance of the differences (Sec. \ref{sub:delta_crit}), we consider errors on different spectral features and perturb the model quantities accordingly, for a range of SNRs (Sec. \ref{sub:err_simul}).

\subsection{SFH relative extension and age resolution}\label{sub:dagen}
For any given star-formation history (SFH) defined by ${\rm SFR}(t)=\frac{dM_{*,{\rm formed}}(t)}{dt}$, we can define its duration $\Delta {\rm age}_{10-90}$ as the time elapsed between the epochs by which 10 per cent and 90 per cent, respectively, of the total stellar mass was formed, following the parameterization introduced by \cite{pacifici+2016}.
Note that we refer to the \emph{formed} stellar mass, i.e. to the time integral of the SFH, which may substantially differ from the present stellar mass, due to the mass fraction returned to the ISM. 
More formally we have:
\begin{equation}
\Delta {\rm age}_{10-90}= {\rm age}_{10} - {\rm age}_{90}
\end{equation}
whereby for $f=10,\,90$, ${\rm age}_f$ is defined by
\begin{equation}\label{eq:agef}
\int_{{\rm age_{Univ}}-{\rm age}_{0}}^{{\rm age_{Univ}}-{\rm age}_f} dt\, {\rm SFR}(t) = \frac{f}{100} M_{*,{\rm total}}
\end{equation}
with ${\rm age}_0$ as the age of the oldest generation of stars (i.e. the look-back time corresponding to the beginning of the SFH), ${\rm age_{Univ}}$ is the age of the Universe today, and 
\begin{equation}
    M_{*,{\rm total}} \equiv \int_{{\rm age_{Univ}}-{\rm age}_{0}}^{\rm age_{Univ}} dt\,{\rm SFR}(t)
\end{equation}

If we normalise $\Delta {\rm age}_{10-90}$ by the median stellar age, ${\rm age}_{50}$ (obtained by plugging $f=50$ into equation \ref{eq:agef}), we obtain what we dub as relative extension of the SFH (or \emph{normalized} $\Delta {\rm age}_{10-90}$):
\begin{equation}\label{eq:dagen}
    \frac{\Delta {\rm age}_{10-90}}{{\rm age}_{50}} \equiv \Delta {\rm age}_{n}
\end{equation}

These definitions can be applied in a model-independent way to any SFH, irrespective of its mathematical formulation.

For a given set of models at fixed ${\rm age}_{50}$, we define as ``age resolution'' the minimum \dagen~resulting in spectra that are significantly different (in a statistical sense) from those having \dagen$\simeq 0$.

\subsection{Library of model SFHs}\label{sub:library}
For the purposes of our analysis we restrict the possible SFHs to the simple analytical form proposed by \cite{Sandage:1986aa}:
\begin{equation}
    {\rm SFR}(t)\propto \frac{t-t_0}{\tau^2}\,{\rm e}^{-\frac{(t-t_0)^2}{2\tau^2}},~ t>t_0
\end{equation}
where $t_0$ is the beginning of the SFH and $\tau$ is a free parameter. The SFR increases quasi-linearly for $t-t_0 < \tau $, reaches a maximum at $t=t_0+\tau$, and eventually declines as a Gaussian of width $\tau$ at later times. No secondary bursts nor any kind of stochasticity is implemented, so that the spectral response to variations in \dagen~is as smooth and regular as possible. In order to avoid metallicity spread or dust attenuation to impact on the ability to resolve small \dagen, we consider only dust-free models and fixed metallicity throughout the SFH. We do expect, however, the metallicity to impact on the age resolution, due to the different prominence of metal absorption features, whose intensity secondarily depends on age as well. Therefore we separately consider models at five different metallicities, namely the native metallicities of the \citetalias{BC03} (v.2016-Padova 1994) SSPs: $2\cdot10^{-2} Z_\odot, 2\cdot10^{-1} Z_\odot, 10^{-1} Z_\odot,  Z_\odot$, and $2.5 Z_\odot$, where $Z_\odot \equiv 0.02$.

Here is how we generate our model library. \emph{i)} For each of the five metallicities, we generate one million models for a variety of Sandage SFHs spanning a broad range of $t_0$ and $\tau$. In particular, we cover the $t_0 - (\tau/t_0)$ space between $10^6$ and $1.4\cdot10^{10}$ years in $t_0$ and between $1/50$ and $2$ in $\tau/t_0$, with a uniform distribution in the logarithm of both quantities. For each model, given $t_0$ and $\tau$, we compute the relevant ${\rm age}_f$, i.e. ${\rm age}_{50}$, ${\rm age}_{10}$, and ${\rm age}_{90}$, and the relative time duration of the SFH, \dagen, as defined in Eq. \ref{eq:dagen}. \emph{ii)} For each model, we compute the spectrum and extract the broad-band photometry in the five SDSS passbands ($u$, $g$, $r$, $i$, and $z$, \citealt{1996AJ....111.1748F}) and the optimal set of absorption indices adopted by \cite{gallazzi+05} for age and metallicity analysis, namely $\rm D4000_n$, ${\rm H}\beta$, ${\rm H}{\delta}_{\rm A}+{\rm H}{\gamma}_{\rm A}$, $\rm [Mg_2Fe]$, and $\rm [MgFe]^\prime$.
The indices are evaluated taking into account an effective velocity dispersion broadening of $\sigma=200\,{\rm km\,s}^{-1}$.
\emph{iii)}~Finally, for each metallicity, we bin the corresponding one million models according to their $\rm age_{50}$ in logarithmic bins of 0.05 dex width, from $10^{6}$ to $10^{10.15}$ yrs. Within each bin we identify the reference model as the one with the minimum \dagen, which, in practice, is an SSP of age equal to $\rm age_{50}$.


\subsection{Simulation of observations at different SNR}\label{sub:err_simul}
The ability to distinguish between different models as a function of \dagen~ depends on the signal-to-noise ratio (SNR) at which the relevant spectral features (indices, broad-band colours, individual spectral pixels) are observed. A relevant part of our experiment is to simulate the statistical uncertainties that affect these features and mock realistic distributions. 

As for the photometric quantities, we assume the typical error budget for SDSS galaxies that are part of the main galaxy sample \citep{strauss_etal02}. As documented in \href{https://classic.sdss.org/dr7/algorithms/fluxcal.php}{https://classic.sdss.org/dr7/algorithms/fluxcal.php} and in \cite{Padmanabhan+2008}, relative calibration errors are of the order of 1--2 per cent for all bands. Photometric random errors are 
 of the same order of magnitude for the $g$, $r$, and $i$ band and roughly a factor two larger for the $u$ and $z$ bands (as per SDSS database on \href{https://skyserver.sdss.org/CasJobs}{https://skyserver.sdss.org/CasJobs}). Since the systematic calibration error is a substantial part of the error budget, we assume the following \emph{uncorrelated} errors, independent of the SNR of the spectroscopic observations: 
 $u_{\rm err}=0.05\,\rm mag$, 
$g_{\rm err}=0.03\,\rm mag$, 
$r_{\rm err}=0.03\,\rm mag$, 
$i_{\rm err}=0.03\,\rm mag$, 
$z_{\rm err}=0.05\,\rm mag$. 

For the spectroscopic observations, we refer to the median SNR per \AA~in a narrow band at $5500-5550$\AA. For a given spectrum, this defines the level of r.m.s. noise per \AA~in that region. We extend such r.m.s. noise to be constant over the entire spectral range. As for the spectral indices, perturbed measurements are obtained by perturbing the index value on the original spectrum using the error computed from the standard propagation theory given the noise spectrum. Where full spectral information is used, the spectrum is perturbed pixel by pixel, by randomly drawing from a Gaussian deviate with $\sigma$ equal to the r.m.s. noise. Note that we do not take into account any correlation among pixels, which may originate, e.g., from spectro-photometric calibration uncertainties. This has a strong impact on the age resolution achievable from un-rectified spectra (see Sec. \ref{sec:results_full_spec}).

\subsection{Statistical determination of the age resolution}\label{sub:delta_crit}
At any given SNR and for each metallicity, we consider narrow bins in $\rm age_{50}$ and the corresponding reference model, as described in Sec. \ref{sub:library}.

For any other model $i$ in the bin, we define the quantity $\delta_i^2$ to measure its deviation with respect to the reference model:
\begin{equation}\label{eq:deltasq}
    \delta^2_i=\sum_{j=1}^{N_{\rm obs}} \frac{(\mathit{obs}_{i,j}-\mathit{obs}_{{\rm ref},j})^2}{\sigma_{i,j}^2}
\end{equation}
where $\{\mathit{obs}_{i,j}\}_{j=1,N_{\rm obs}}$ are the observables measured on the $i$-th model and $\sigma_{i,j}$ the corresponding errors, while  $\{\mathit{obs}_{\rm ref,j}\}_{j=1,N_{\rm obs}}$ are the observables measured on the reference model. As mentioned in the introduction, in this work we will use different sets of observables, including spectral indices, broad-band colours and fluxes in individual spectral pixels.

\begin{figure}
	\includegraphics[width=\columnwidth]{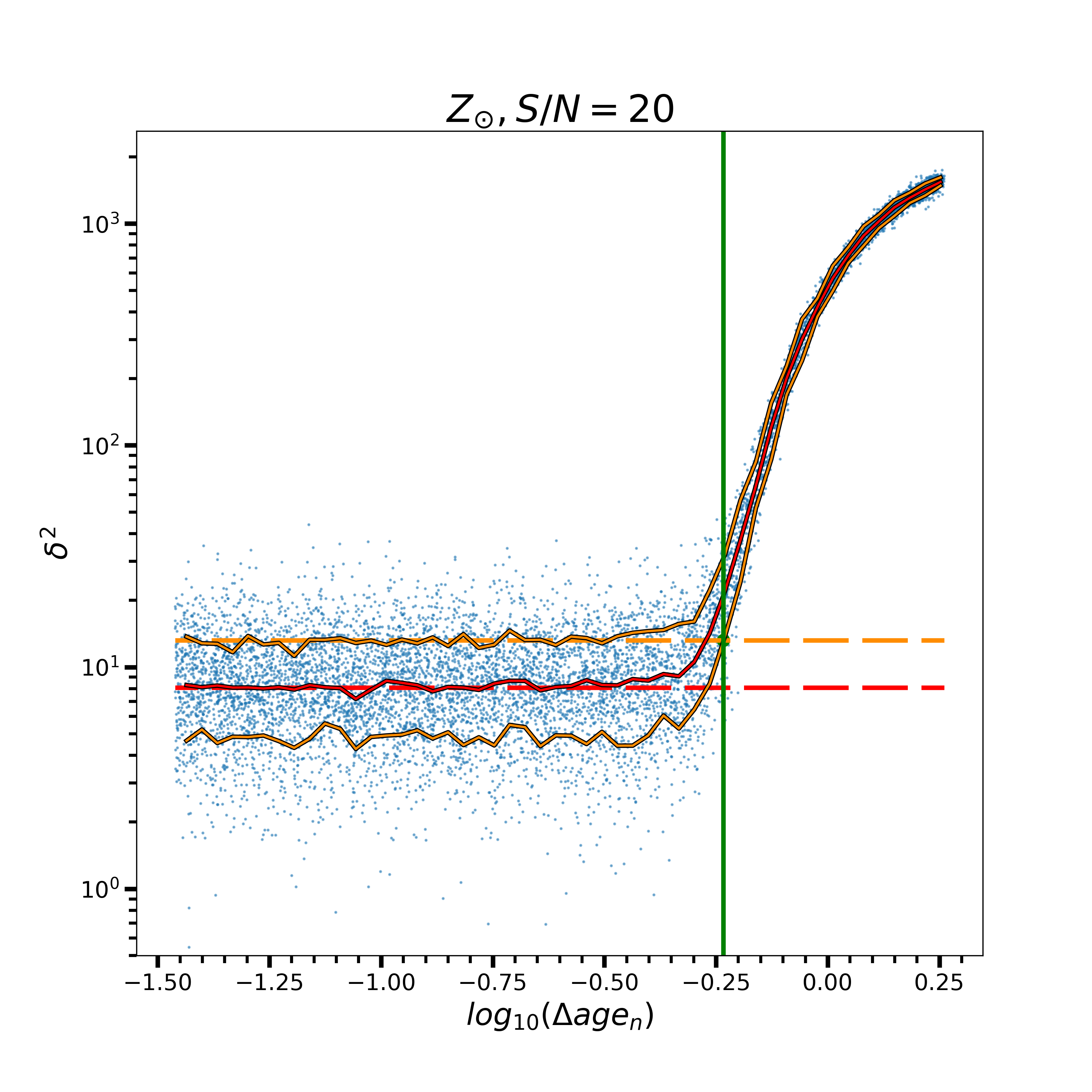}
    \caption{Example of a distribution of the statistical estimator for spectral differences, $\delta^2$, as a function of the relative SFH duration \dagen, for a narrow age bin of our model library. Specifically, here we report the data for the spectral indices and colours as observable set, in the bin $10^{9.7}<{\rm age}_{50}/{\rm yr}<10^{9.75}$, $Z_*=\rm Z_\odot$, assuming ${\rm SNR}=20$\invAA. The blue dots represent the $\delta_i^2$ for each perturbed model $i$. The solid red and orange lines represent the running median and the running $16^{\rm th}$ and  $84^{\rm th}$ percentiles, respectively. The horizontal long-dashed lines mark the median (red) and the $84^{\rm th}$ percentile (orange) in the reference range (i.e. $\log(\Delta{\rm Age_n})<-1$). The vertical green line marks the value of \dagenmin, corresponding to the \dagen~where the running $16^{\rm th}$ percentile equals the $84^{\rm th}$ percentile in the reference range.}
    \label{fig:delta_dagen_example}
\end{figure}

Fig. \ref{fig:delta_dagen_example} shows the typical distribution of $\delta^2$ as a function of \dagen~for a narrow age$_{50}$ bin at fixed metallicity, considering a given set of observables and given SNR.
At low \dagen~the distribution in $\delta^2$ is remarkably constant and consistent with pure noise fluctuations. As we increase \dagen~ to exceed values $\gtrsim 10^{-0.4}$, $\delta^2$ rapidly increases, indicating that models with such a large \dagen~are significantly different from the reference model of approximately null SFH duration. In order to quantitatively determine the inflection point in a statistically robust way, we consider the rolling percentiles of the $\delta^2$ distribution as a function of \dagen. We define the minimum \dagen~that allows to reliably detect a difference in spectral features, \dagenmin, as the point where the rolling 16-th percentile of the $\delta^2$ distribution (lower orange line in Fig. \ref{fig:delta_dagen_example}) crosses the 84-th percentile of the distribution in a reference range of $\simeq 0$ duration (dashed orange horizontal line). As a reference range we choose \dagen$<10^{-1}$, as we have checked that trends are always flat in that range\footnote{Note that the typical age sampling of the adopted SSPs, ranging between a few 0.001 and 0.05 dex, allows us to densely sample short duration SFHs, well below the ``short duration limit'' of \dagen$<10^{-1}$.} for any combination of age, metallicity, spectral features and SNR. The adopted operational definition of \dagenmin~is essentially a 2-$\sigma$ criterion and should be regarded as an optimal compromise between a good statistical robustness and not being too conservative in the detection of spectral differences.


\section{Results I: age resolution from spectral indices and broad-band colours}\label{sec:results_indx_phot}
In this section we analyze the age resolution that is possible to achieve under the ideal conditions described above, based on broad-band colours and a set of selected spectral absorption features.

\begin{figure*}
	\includegraphics[width=\textwidth]{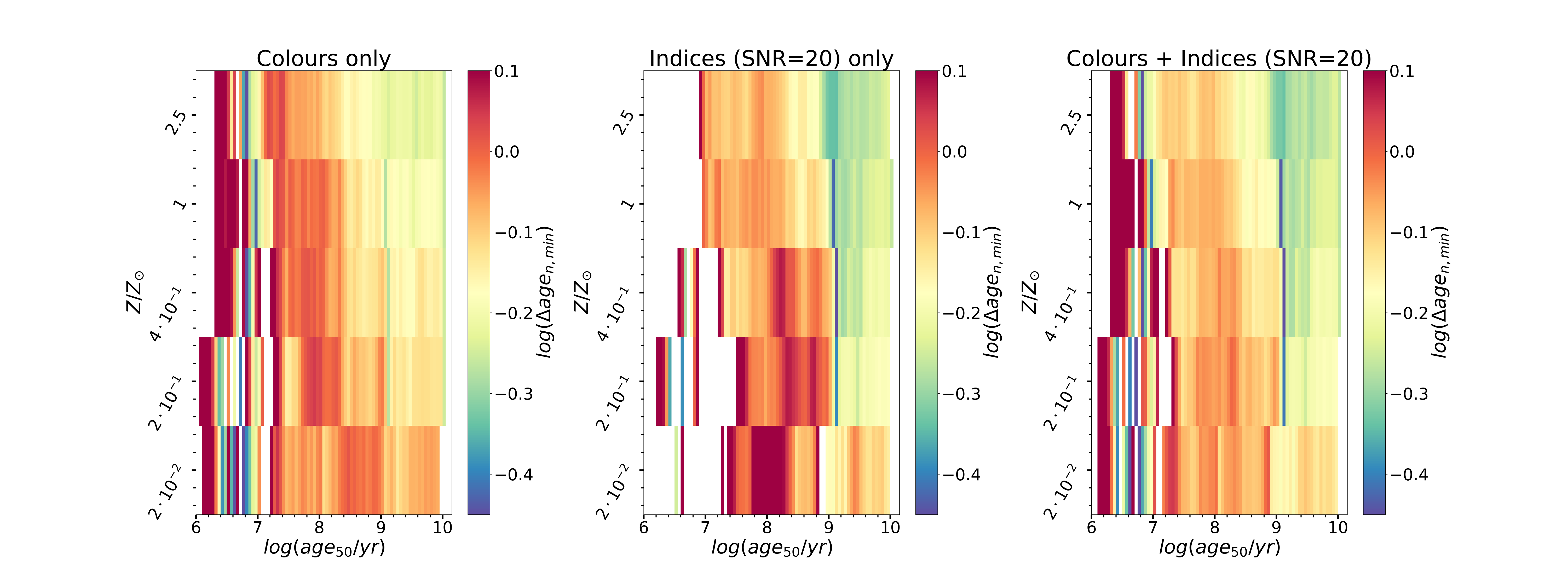}
    \caption{Map of age resolution \dagenmin~as a function of age ($x$-axis) and metallicity ($y$-axis), as determined by broad-band SDSS colours alone (\emph{left panel}), by spectral indices alone (\emph{central panel}), and by broad-band colours and spectral indices jointly (\emph{right panel}), assuming a spectral ${\rm SNR}=20$\invAA. Blank bins denote age/metallicity bins in which the procedure for the computation of \dagenmin~failed, due to the chaotic evolution of spectral features around those ages.
    }
    \label{fig:ageres_maps}
\end{figure*}

First of all, in the left panel of Fig. \ref{fig:ageres_maps} we show the age resolution as determined by variations in the five SDSS passbands, i.e. by variations in the low-frequency shape of the optical SED, quantified by the four independent colours $u-r$, $g-r$, $r-i$ and $r-z$. For the 5 million models of our library, binned in 0.05-dex wide bins of median age (along the $x$ axis) and according to the five fixed metallicities (along the $y$ axis), the age resolution \dagenmin~is represented by the colours, according to the colourbar. Across the age-metallicity plane we see that the typical age resolution achievable with broad-band colours alone is $-0.25\lesssim \log~$\dagenmin$\lesssim -0.05$, corresponding to $0.55\lesssim $\dagenmin$\lesssim 0.90$. From this map we can notice two general trends, as a function of age and of metallicity. \emph{i}) At given age, the age resolution improves going from low to high metallicity. This is most likely a consequence of the dependence on photospheric temperature (hence on age) of the metal absorption blanketing, which, in turn, translates into colour variations. At higher metallicities such variations are stronger and therefore more promptly detected.  \emph{ii}) At fixed metallicity, the general trend is for better age resolution at ages around and/or above $\simeq 1$~Gyr. For $Z_*\gtrsim 0.2\,\rm Z_\odot$ we observe a minimum in \dagenmin~at median ages around $\simeq 1$~Gyr. The age resolution worsens systematically below $\simeq 1$~Gyr, down to a few $10^7$~yr of age. At even younger ages the trends become less clear and we can also note several bins where \dagenmin~is undefined (blank bins). As we will show below with absorption indices (see Sec. \ref{sub:disc_spec_feat} and Fig. \ref{fig:index_time_deriv}), these young ages present very fast and somehow chaotic spectral evolution, so that the procedure for the determination of \dagenmin~presented in Sec. \ref{sub:delta_crit} fails to converge because of the non-monotonic behaviour of the median trend of $\delta^2($\dagen$)$.

The central panel of Fig. \ref{fig:ageres_maps} displays the trends for the age resolution \dagenmin~based on the optimal set of five spectral absorption indices alone, namely $\rm D4000_n$, ${\rm H}\beta$, ${\rm H}\delta_{\rm A}+{\rm H}\gamma_{\rm A}$, $\rm [Mg_2Fe]$, and $\rm [MgFe]^\prime$. In this case we have adopted a SNR of 20 per \AA, which we can assume as representative of (the good quality tail of) large extragalactic spectroscopic surveys for stellar populations studies, such as the SDSS \citep[see][]{gallazzi+05}, LEGA-C \citep{lega_c_DR3} or WEAVE-StePS \citep{Iovino+2023}. While trends are qualitatively similar to the case of photometry alone, there is a general decrease of \dagenmin~over most of the age-metallicity plane. The improvement in age resolution is particularly large both for median ages above 1 Gyr and for metallicities above $0.2~\rm Z_\odot$, with a typical decrease of \dagenmin~by $\simeq 0.1$~dex, reaching down to \dagenmin$\lesssim 0.5$. In the regime of low age (${\rm age}_{50}<1$~Gyr) and low metallicity ($Z_* <0.4\, {\rm Z_\odot}$), the spectral indices as measured on a SNR$=20$\invAA~spectrum provide an age resolution that is comparable to or even worse than the one provided by optical photometry alone. This is not particularly surprising if we consider that in such a range the strength of the absorption features is significantly reduced. As we will see below, the superior information included in the spectra emerges as soon as one pushes to higher SNR. 

In the right panel of Fig. \ref{fig:ageres_maps} we show the age resolution resulting from combining both photometric and spectroscopic constraints. As an obvious consequence of the trends noted before, adding the photometric constraints has a significant impact only in the range of low-ages/low-metallicities. We obtain an age resolution of $0.4\lesssim$\dagenmin$\lesssim 0.7$ over most of the age-metallicity plane, except for ages less than a few 10-million years, where the resolution degrades substantially and/or \dagenmin~becomes even impossible to measure. This substantial uniformity across approximately three orders of magnitude in age and two orders of magnitudes in metallicity is indeed remarkable. However, it is noteworthy that at all metallicities the age range around 1 Gyr is the spot where the best resolution can be achieved.

\subsection{Impact of spectroscopic signal-to-noise ratio on the time resolution}
Next we want to explore how the SNR of the spectra impacts the age resolution in the ideal cases as set up for our experiment, and, hence, how far it is desirable to push for deep spectroscopy in order to make a detailed reconstruction of a galaxy's SFH. To this goal, we repeat the same procedure for determining \dagenmin, but now using spectral indices perturbed according to different SNRs per \AA, namely: 5, 10, 20, 50, 100, 200 and 500.

\begin{figure*}
	\includegraphics[width=\textwidth]{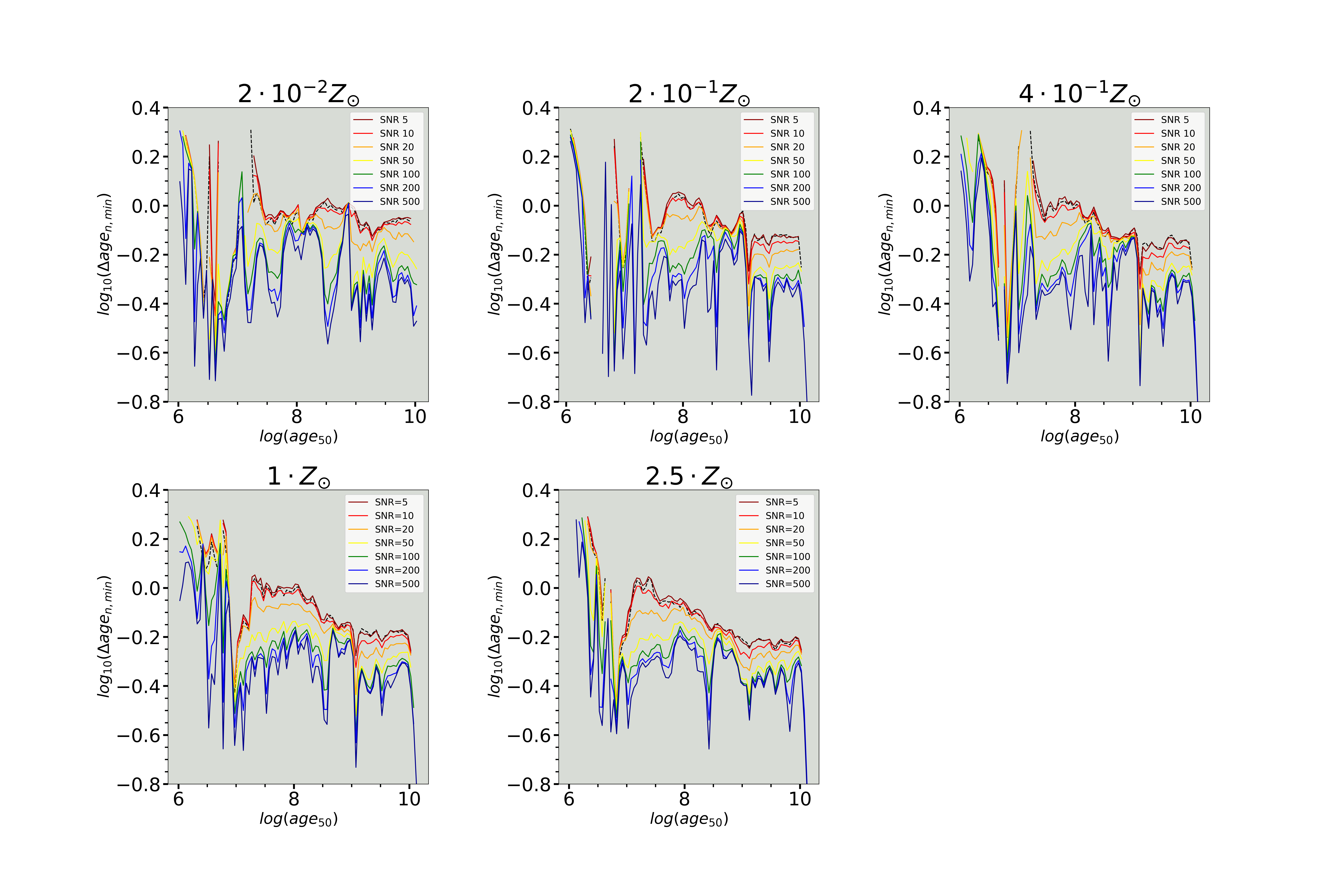}
    \caption{Age resolution \dagenmin~as a function of age ($x$-axis) as determined from photometry and spectral indices at different SNR per \AA~(lines in different colours according to the legend), at five different fixed metallicities (different panels). For a reference, the black dashed lines display the age resolution from photometry alone. Remarkably, the gain in age resolution in going from 50-100\invAA~in SNR to 500\invAA~is very marginal at most ages and metallicities. Missing segments denote age bins in which the procedure for the computation of \dagenmin~failed, due to the chaotic evolution of spectral features around those ages.}
    \label{fig:ageres_snr_all}
\end{figure*}

The five panels of Fig. \ref{fig:ageres_snr_all} display the age resolution as a function of age at the five fixed metallicities of our spectral libraries, respectively. For each $\rm age_{50}$, \dagenmin~is estimated combining the photometric and the spectroscopic constraints. Each coloured line is computed assuming one of the seven different spectral SNRs, from dark red to purple, according to the legend. The black dashed lines display \dagenmin~as a function of $\log \rm age_{50}$ as obtained from photometric constraints alone, thus corresponding to a null SNR in the spectrum. As expected, the age resolution improves at higher SNR. However, the quantitative variations are of particular interest. At ${\rm SNR} \lesssim 10$\invAA, spectroscopy does not provide any particular advantage over photometry alone. Note that this statement is strictly valid in the context of our idealized experiment where median age and metallicity are fixed and no dust is included. In a real case, we do expect a substantial difference in the accuracy of the age and metallicity determinations moving from 0 to 10 in SNR per \AA, as the spectroscopic information (contrary to broad band photometry alone) can substantially reduce the degeneracy among different physical parameters.

A substantial gain in age resolution occurs when increasing the spectral SNR from 10\invAA~up to 50-100\invAA. Depending on the median age and metallicity, the gain ranges between $\simeq 0.05$~dex up to $\lesssim 0.3$~dex (almost a factor 2). Quite remarkably, with the only exception of a few specific and limited ranges in age and metallicity, increasing the SNR beyond 100\invAA~does not result in substantial improvements in the age resolution. This is also due to intrinsic scatter in the spectral properties induced by the finite width of 0.05 dex for the bins in $\rm age_{50}$, which is meant to reproduce the minimal uncertainty that can be obtained with an optimal dataset.

Not all age ranges appear to benefit from high-SNR spectroscopy in the same amount. 
The age range around $10^{7.5}$-$10^{8}$~yr is the most impacted one at all metallicities.
At $Z_*\gtrsim \rm Z_\odot$ the age resolution improves by 0.1 dex up to $\lesssim 0.3$~dex over the whole age range between $\simeq 10^7$ and $~\simeq 10^{8.5}$~yr. A more limited improvement by $\simeq 0.1$~dex is observed over the same SNR range for ages $\gtrsim 1$~Gyr. For these metal rich stellar populations, the overall effect of moving from low to high SNR is to make the age resolution more uniform around \dagenmin$\simeq 0.5$ by most effectively improving the young age regime.
At metallicities lower than solar, \dagenmin~is characterized by stronger variations with $\rm age_{50}$, and the amplitude of the trends with SNR appear also to depend on the $\rm age_{50}$ very strongly. At sub-solar metallicity, the age ranges that benefit most from high-SNR spectroscopy are those between 10 and 100 Myr, and above $\simeq 1$~Gyr, respectively.

\section{Results II: Age resolution with full spectral fitting}\label{sec:results_full_spec}

It is often argued that spectra contain much more and more redundant information than that carried by (a selection of) the classic absorption indices. This would make the full spectral fitting much more effective than the fitting of absorption indices in recovering stellar population properties, especially when using low-SNR spectra. Actually, this is one of the reasons why full-spectral-fitting codes have become so popular. 

In a similar way as done in the previous section with indices and photometry, in this section we explore the age resolution that can be attained by taking into account the spectral information pixel-by-pixel. 
At any given metallicity and for each bin in $\rm age_{50}$, we quantify the deviation of each model $i$ from the reference one of null duration, by means of the $\delta^2_i$ parameter, defined in Equation \ref{eq:deltasq}. In this case $\{\mathit{obs}_{\rm ref,j}\}_{j=1,N_{\rm obs}}$ are the flux densities $f_\lambda$ of each of the $N_{\rm obs}$ individual pixels in a given wavelength range. In a first approach we concentrate only on the ``high-frequency'' information carried by the absorption features rather than on the ``low-frequency'' information carried by the spectral shape\footnote{Note that, in general, the flux calibration accuracy of typical extragalactic spectra is not competitive with photometric measurements obtained from imaging observations in order to provide accurate constraints on the broad spectral shape of a galaxy's SED. In fact, the possibility of ``adjusting'' the spectral shape to match the SPS templates by means of additive or multiplicative polynomials is foreseen by some codes, most notably by \textsc{pPXF} \citep[][]{Cappellari_pPXF}.}. Therefore, we \emph{rectify} each spectrum before performing the analysis. Specifically, in a first pass we normalize each spectrum so to have unity flux density at $5500$~\AA. In a second pass, we compute a smoothed version of the spectrum by means of a running median filter, adopting a top-hat $450$\AA-wide kernel. The full-resolution normalized spectrum is finally divided by the smoothed one to obtain the rectified (continuum-normalized) spectrum, which is used for the relative comparisons, the computation of $\delta^2_i$ and, finally, for determining \dagenmin.

\begin{figure*}
 \centering
    \includegraphics[width=\textwidth]{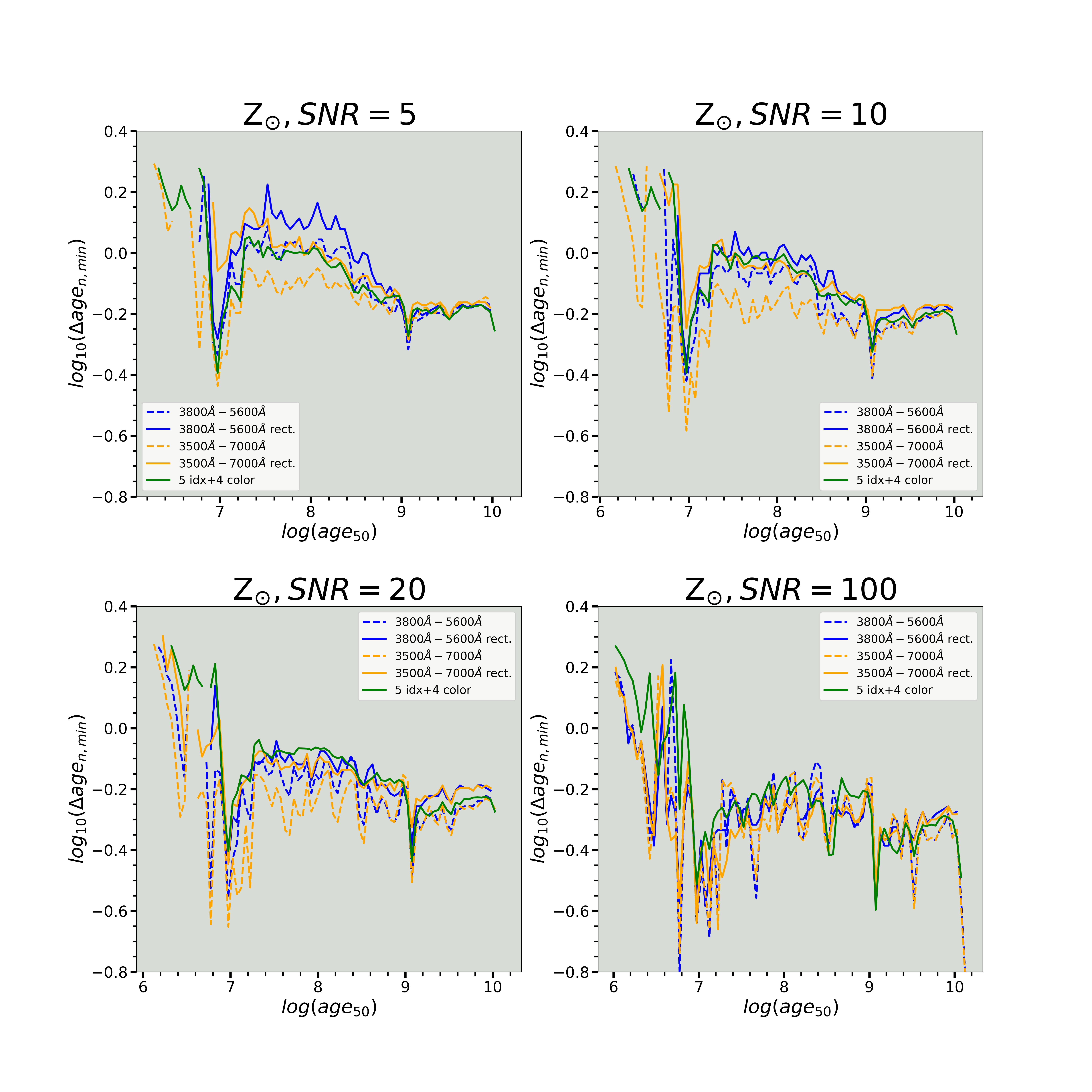}
    \caption{Comparison of the age resolution \dagenmin~achievable based on indices and colours (\emph{green line}) with that obtained from full spectral fitting, using two different wavelength ranges (narrow range in \emph{blue} and broad range in \emph{orange}). For the full spectral fitting we considered two different approaches, one using rectified spectra (i.e. discarding the information about the low-frequency SED shape, represented with solid lines) and the other using the spectra as they are (un-rectified, represented by dashed lines). For illustration purposes we report the results for solar metallicity only and for four different SNR per \AA, as reported in each panel. The analogous plot for $Z=0.2\,Z_\odot$ is reported in Appendix \ref{app:fullspec_20perc}, Fig. \ref{fig:FSF_20pecentZsolar}. Missing segments denote age bins in which the procedure for the computation of \dagenmin~failed, due to the chaotic evolution of spectral features around those ages.}
    \label{fig:FSF_Zsolar}
\end{figure*}

The adoption of the full-spectral-fitting approach on rectified spectra (hence neglecting the broad-band SED shape) reproduces the age resolution and the trends obtained using indices and photometry, both qualitatively and quantitatively, within a few 0.01 dex. In Fig. \ref{fig:FSF_Zsolar} we focus on the models at solar metallicity and present a detailed comparison between the \dagenmin~obtained based on indices and photometry (\emph{green lines}) on the one hand, and full spectral fitting, in two different wavelength ranges (\emph{blue and orange solid lines}) respectively, on the other hand. Results are reported for four different SNR in the spectra, namely 5, 10, 20 and 100\invAA. This SNR was adopted to perturb both the spectra and the indices consistently. We consider a narrow wavelength range 3800-5600\AA~(\emph{blue lines}), which roughly corresponds to the range covered by the absorption indices, and a wide range 3500-7000\AA~(\emph{orange lines}), which encompasses the full optical extent.

As we can see from the figure, at SNR around 10--20\invAA~the age resolution achievable with the three different sets of constraints is very similar overall across the ages from $10^7$ to $10^{10}$ years. If we consider the old age range, $\rm age_{50} \gtrsim 10^{9}\rm yr$, we notice that indices+colours perform significantly better than the full spectrum at SNR of 20\invAA~and 10\invAA, while indices+colours still perform marginally better both at very low (5\invAA) and very high (100\invAA) SNR. In the young age regime ($10^7 \rm yr \lesssim \rm age_{50} \lesssim 10^{9}\rm yr$) indices+colours are the most effective constraints at SNR$\leq 10$\invAA, yet with a performance very similar to the full spectrum in the broad wavelength range. In particular, at very low SNR the narrow spectral range performs about 0.1 dex worse than both the broad range and the indices+colours. At SNR $\geq 20$\invAA~the full spectrum appears to perform increasingly better as SNR increases, although typically by no more than a few 0.01 dex relative to indices+colours, with no much difference between the narrow and the broad wavelength range.

Very similar results are obtained at subsolar metallicity (see Appendix \ref{app:fullspec_20perc}).

One could argue that rectified spectra miss the significant piece of information connected to the spectral shape, which is instead considered in the indices+colours analysis. Indeed, as already noted in the previous section, by looking at Fig. \ref{fig:ageres_maps} and \ref{fig:ageres_snr_all} we can see that the information about the spectral shape carried by the photometry is only relevant for ages below 1 Gyr and relatively low metallicities, and in any case only at SNR$\lesssim 20$, if realistic although ideal photometric uncertainties are assumed.

We further investigate this issue by repeating the analysis of this section and applying our statistical estimator $\delta^2$ to un-rectified spectra, which preserve the information about the SED shape. The results are reported as dashed lines in Fig. \ref{fig:FSF_Zsolar} for $Z=\rm Z_\odot$ (and Fig. \ref{fig:FSF_20pecentZsolar} for $Z=0.2\,\rm Z_\odot$).
Using un-rectified spectra improves the age resolution achievable with full spectral fitting especially at low SNR, in general, also with respect to indices+colours. Differences up to $0.15$\,dex (or more, at some specific ages) are reached at SNR$=5$, while the typical differences decrease to less than 0.1 dex when the SNR reaches 20 or more. The adopted spectral range is also relevant, with the wide $3500-7000$\AA~range providing much better resolution than the narrower $3800-5600$\AA~range.
The improvement due to the adoption of un-rectified spectra is stronger for young ages (below $\approx 1$\,Gyr), while it becomes almost negligible for older ages. This is consistent with the results of Sec. \ref{sec:results_indx_phot}, where we showed that the age resolution is more sensitive to broad-band colours in the low-age (and low-metallicity) regime, while at old ages (and high metallicity) spectral features, as those quantified by the indices, become the dominant constraints.

While at face value these results strongly support the use of un-rectified spectra, it is important to realize that in this experiment flux calibration errors are not properly taken into account. Each model spectrum is perturbed solely based on the noise of individual pixels, which are assumed to be uncorrelated. In reality, spectra suffer by errors in the spectro-photometric calibration that can be hardly reduced below a few-percent level and affect pixels in a systematically correlated way. As a consequence, the error on the spectral shape (e.g. quantified by broad-band colours obtained as synthetic photometry via spectral integration) are severely under-estimated in a simple pixel-by-pixel full spectral fitting procedure.
In fact, broad bands like the SDSS $ugriz$ encompass a number of spectral pixels of the order of $10^3$. Thus, the SNR of the broad-band flux measurements is a factor of the order of $\sqrt{10^3}\approx 30$ larger that the spectral SNR per \AA. Already for a SNR of 5\invAA, this corresponds to errors of less than 0.01 mag on the broad-band photometry, which is hardly plausible. Even smaller and totally unrealistic errors of  $\lesssim 0.001$~mag would correspond to larger SNR.
In conclusion, the results from the full spectral fitting with un-rectified spectra must be regarded as formal limits for the age resolution, with realistic estimates being in between the results from un-rectified and rectified spectra. In any case, Fig. \ref{fig:FSF_Zsolar} and \ref{fig:FSF_20pecentZsolar} show that for SNR$\gtrsim 20$\invAA~the difference in age resolution obtained with the different methods substantially align, as expected if most of the information is carried by spectral features or if the SED shape provides only redundant information.

We can thus conclude that a full spectral fitting approach does not substantially improve the age resolution over the indices+colours approach, in general. In fact, the relevant information encoded in absorption features appears to be equally well captured by the indices and by the rectified spectra, with a small preference for indices at very low SNR. The age resolution benefits from accurate spectral shape information (which can be provided either by broad-band colours or by un-rectified spectra) especially at young ages (and metallicities) and low SNR, i.e. in the regime in which the absorption features are the weakest and hardest to measure.

We add here a final remark about the possible effect of spectral resolution. While index measurements were corrected to a Doppler velocity broadening of $\sigma=200~\rm km~s^{-1}$, for computational reasons we have run our simulations on full spectra at the native resolution of the BC03+MILES models, corresponding to 2.5\AA~FWHM. This corresponds to a Doppler broadening $\sigma\simeq 65~\rm km~s^{-1}$ at $\sim 4500-5000$\AA~(i.e. in the middle of the wavelength range)\footnote{Since MILES models have a constant resolution in $\lambda$, the resolution in terms of velocity dispersion ranges between a maximum of $\sim 85~\rm km~s^{-1}$ at the blue end at 3500--3800\AA~and a minimum of $45~\rm km~s^{-1}$ ($57~\rm km~s^{-1}$) at the red end at 7000\AA~(5600\AA).}. While a higher spectral resolution is reasonably expected to enable higher age resolution by resolving finer and much more numerous spectral features (provided that a high enough SNR is reached), most studied galaxies have velocity dispersion similar to or significantly higher than this. Indeed, the performance of the full spectral fitting quoted in this section may well be overoptimistic in a realistic case of galaxies with velocity dispersion of the order of 100 to $\gtrsim 300~\rm km~s^{-1}$. On the other hand, it is worth noting that absorption indices are only mildly sensitive to the velocity broadening as long as $300-400~\rm km~s^{-1}$ are not exceeded (as we verified on our models), as a consequence of the width of the central- and side-bands extending to a few tens of \AA.

\section{Discussion}\label{sec:discussion}
\subsection{Validity and limitations of our results}\label{sub:limitations}
Before discussing the implications of our results and the possible physical interpretations of the observed trends, it is worth recalling the framework and the limitations in which they have been derived. In this paper we wanted to work out the theoretical limits to our ability to detect a finite duration of a SFH relative to an instantaneous (zero-duration) burst. For this reason, we have considered only idealized cases (smooth SFH, single metallicity, no dust) and basic statistical estimators ($\delta^2$, as defined in Sec. \ref{sub:delta_crit}).

The treatment of spectral noise is also very much simplified. First of all, we assume Gaussian noise, as done in all spectral fitting codes. Moreover, we assume constant noise per \AA, which is clearly not realistic, as it does not take into account the wavelength dependence of the spectrograph and detector efficiency and of the atmospheric transmission. These effects are very much specific to instruments and galaxy redshift, so that it is virtually impossible to cover all possibilities. However, over the spectral range covered by our analysis, the flux density of a galaxy varies by a factor of a few \citep[see, e.g., figure 9 and 10 of][]{BC03} and the same holds true also for the level of noise and transmission as long as we work inside the visible window. For these reasons, the levels of SNR quoted in our results should be taken as just indicative in absolute terms, while retaining their validity in relative terms and as long as trends are concerned.

Also concerning the treatment of noise, it is worth noting that models themselves are affected by the noise inherited by the stellar templates adopted (in our case the MILES library). This noise term is a function of the age and metallicity and it is technically very hard to deal with. In fact, empirical stellar spectra have a SNR of 150 at most \citep{Sanchez-Blazquez:2006aa}, and this might be one of the reasons why the age resolution saturates at SNR~$\gtrsim100$\invAA.

In this work we have deliberately considered only the visible wavelength range, in which most of the stellar-population and SFH analysis has been focusing in the past. We thus excluded the ultra-violet (UV, below $\sim 3500$\AA) and the near infrared (NIR, above $\sim 1\mu$m), which are typically also affected by larger SPS model uncertainties. It is worth noting, however, that the UV coverage may indeed improve the age resolution at ages below $\sim1$~Gyr, thanks to the strong impact of hot short-lived stars in this wavelength range. In fact, recent works by, e.g., \cite{costantin+2019} and \cite{salvador-rusinol+2020} showed that complementing the visible spectra with the UV coverage enables to capture minor rejuvenation episodes (age $\lesssim 1$~Gyr) on the top of relatively old stellar populations.

Finally, it is important to realize that the minimum resolvable relative duration of the SFH, \dagenmin, is not representative of the accuracy that can be reached in the determination of the relative duration of the SFH \dagen~in realistic cases. Just as an illustration, in appendix \ref{app:dagen_recovery} we report the results about the accuracy in the retrieval of \dagen, obtained for a large variety of models with complex SFH, chemical enrichment and dust attenuation as in \cite{Zibetti:2017aa}, using the Bayesian marginalization approach as in \cite{gallazzi+05} and \cite{Zibetti:2017aa}, for a representative SNR~of 20\invAA. This analysis shows that in a realistic scenario only \dagen$\gtrsim 1$ (i.e. SFH durations of the order of the median stellar age) can be recovered in an unbiased way, with typical uncertainties around 0.1--0.2 dex. SFH durations significantly shorter than the median stellar age tend to be largely unconstrained and display a posterior PDF that closely mirrors the prior PDF. These results account for the degeneracies between SFH parameters, metallicity and dust attenuation, as obtained from the marginalized posterior PDF.

\subsection{Comparison with the literature}\label{sub:disc_literature}
As already stressed in the introduction, the age resolution attainable via spectral fitting determines the performance limits of the spectral inversion algorithms that aim at reconstructing the SFH or even the full age-metallicity distribution of a composite stellar population. Based on our results, the minimum size of the bins under which a SFH can be sampled vary between $\approx$ 0.3 and 0.7 times the median age, or, if logarithmic sampling is used, between 0.15 and 0.25 dex.

The issue of age resolution in the reconstruction of SFHs has been treated with different approaches in the literature. \cite{Ocvirk+2006} quantified the age resolution in terms of the minimum age separation of two bursts of equal intensity that their \textsc{stecmap} code was able to disentangle. They found a resolution $\Delta \log \rm Age_{Ocvirk}$ varying between $\simeq 0.9$~dex$ (\simeq 1.5$~dex) and $\simeq 0.4$~dex ($\simeq0.6$~dex) as a function of SNR from 20 to 100 per \AA, at fixed (variable) metallicity, with a saturation above SNR$\simeq100$\invAA. In order to convert these numbers in terms of \dagenmin, we consider
\begin{dmath}
    \Delta {\rm age_n}=\frac{{\rm Age_{burst 2}}-{\rm Age_{burst 1}}} {<\rm Age>} = \frac{{\rm <Age>}\times 10^{\Delta \log \rm Age_{Ocvirk}/2}-{\rm <Age>}\times 10^{-\Delta \log \rm Age_{Ocvirk}/2}} {<\rm Age>}\\
 = 10^{\Delta \log \rm Age_{Ocvirk}/2}- 10^{-\Delta \log \rm Age_{Ocvirk}/2}
\end{dmath}.
We thus derive that the resolution reported by \cite{Ocvirk+2006} corresponds to values of \dagenmin~between 2.5 (5) and 1 (1.5), which is actually much more pessimistic than our results imply. This can be understood by considering that \cite{Ocvirk+2006} require the two bursts to be well resolved and correctly dated, while our requirement is much milder, as we only request to distinguish between null and extended duration.

In \textsc{VESPA} \citep[][]{Tojeiro+2007} the choice of age bins is automatically made by the code, which is able to adapt to the SNR of the observations, based on singular value decomposition (SVD) criteria. For data in the SNR range 20\invAA~to 50\invAA~the code is able to assign weights to no more than 5 bins per dex, i.e. with a time resolution of 0.2 dex. This is very much in line with our results that find minimum bin sizes between 0.15 and 0.25 dex (\dagenmin~of $\approx-0.35$).

Other codes, like \textsc{ppxf} \citep{Cappellari_pPXF,Cappellari:2017} and \textsc{starlight} \citep{STARLIGHT} work on a fixed grid of SSPs, therefore do not directly deal with the issue of time resolution. However, since \cite{Cappellari:2017}, it has become obvious that the SSP decomposition output by these codes requires some regularization in order to avoid severe overfitting on small time scales. In fact, as one can see e.g. from Figure 3 in \cite{Cappellari:2023}, even at SNR$>20$\invAA~one can hardly resolve SF episodes shorter than 0.3 dex (corresponding to $\log$~\dagenmin$\simeq -0.3$) if the appropriate regularization is applied.

\subsection{Physical interpretation of the age-resolution trends}\label{sub:disc_physical}
\begin{figure*}
 \centering
    \Large $9.5<\log({\rm age}_{50})<9.55, Z=Z_{\odot}$\par \vspace{-1cm}
    \includegraphics[width=\textwidth]{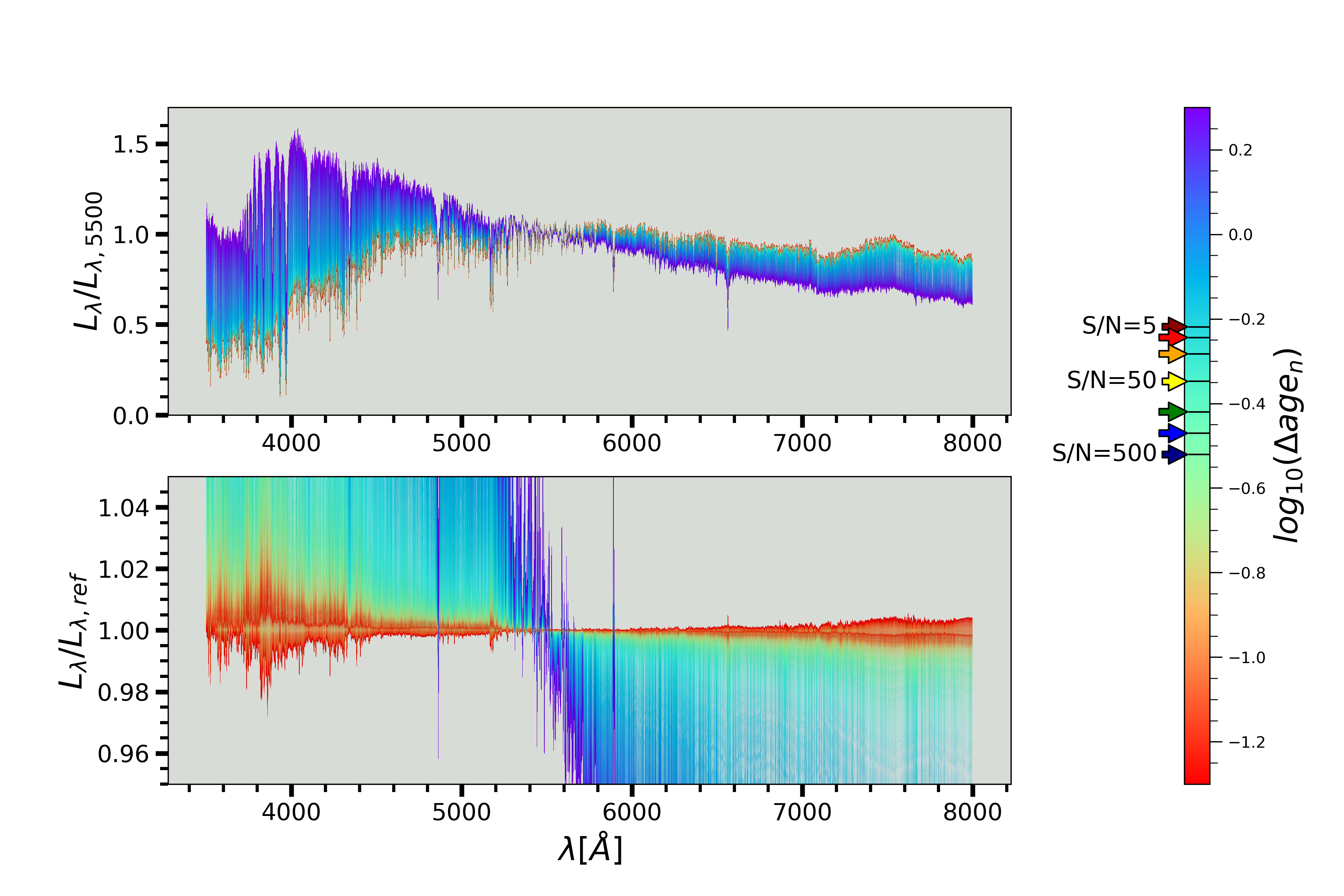}
    \caption{
    \emph{Top panel}: Model spectra in the age bin $9.5 <\log(\rm age_{50}/yr) <9.55 $ and $Z=Z_\odot$. The spectra are normalized between 5450\AA~and 5550\AA~and color-coded according to their \dagen, as shown in the colorbar. \emph{Bottom panel}: ratio between each spectrum and the ``reference'' spectrum, defined as the spectrum of the model with the lowest \dagen-value.
    The painted arrows next to the colorbar represent $\log($\dagenmin$)$ for different signal-to-noise ratios: Dark red: SNR=5\invAA; red: SNR=10\invAA; orange: SNR=20\invAA; yellow: SNR=50\invAA; green: SNR=100\invAA; blue: SNR=200\invAA; dark blue: SNR=500\invAA.}
    \label{fig:stacked_specs}
\end{figure*}

In this section we analyze in greater detail the physical origin of the trends observed in the previous sections. Let us consider a bin in $\rm age_{50}$ and $Z$ in our model library and study how the spectral shape and index strengths change as a function of \dagen. For the sake of clarity and conciseness, we take as a representative example the bin corresponding to $9.5<\log({\rm age}_{50}/{\rm yr})<9.55$ and $Z=Z_{\odot}$, where the spectra display very clearly both Balmer and metal absorptions. As explained in Sec. \ref{sub:delta_crit}, models with very small \dagen~are all very similar to each other and to the SSP of ${\rm age}={\rm age}_{50}$. The effect of increasing the duration of the SFH (i.e. taking models with larger \dagen) is to add both younger and older populations with respect to the median age, ${\rm age}_{50}$. However, these two tails do not average out in the spectrum of the composite stellar population because of the dependence of the mass-to-light ratio $M/L$ on age. Since younger stellar populations have lower $M/L$, increasing \dagen~results in spectra that are increasingly dominated by stellar populations that are younger than ${\rm age}_{50}$, thus, generally speaking, bluer. This effect is clearly illustrated in Fig. \ref{fig:stacked_specs}, where all the spectra of the bin are plotted (after flux normalization at $5500$\AA), colour-coded according to their \dagen. The top panel displays the actual spectra, while the bottom panel displays the ratio of each spectrum over the reference spectrum of virtually null duration (see Sec. \ref{sub:delta_crit}). These plots dramatically illustrate the lack of significant differences in terms of spectral shape for models with \dagen~up to $\approx 10^{-0.5}\simeq 0.3$, which all basically overlap in the top panel. Longer SFH timescales translate into systematically bluer shapes, but systematic deviations as a function of \dagen~are impossible to measure until $\log$~\dagen$\simeq -0.5$ is reached, because of their small amplitude, at $\approx 1\%$ level. The arrows on the colour-bar indicate the age resolution \dagenmin~for different values of SNR, in order to illustrate the ability to pick up spectral differences in different noise regimes.
It is worth noting that part of the scatter for very low values of \dagen$\lesssim 10^{-1}$ is due to the finite (although small) range of ${\rm age}_{50}$ considered in each bin. Knowing the exact value of ${\rm age}_{50}$ would of course enable a more accurate estimate of \dagen, but in reality we have to take into account that these two quantities are fundamentally degenerate, so that it is impossible to estimate one of the two assuming negligible uncertainty on the other one.

\begin{figure*}
 \centering
    \Large $9.5<\log({\rm age}_{50})<9.55, Z=Z_{\odot}$\par \vspace{-1cm}
    \includegraphics[width=\textwidth]{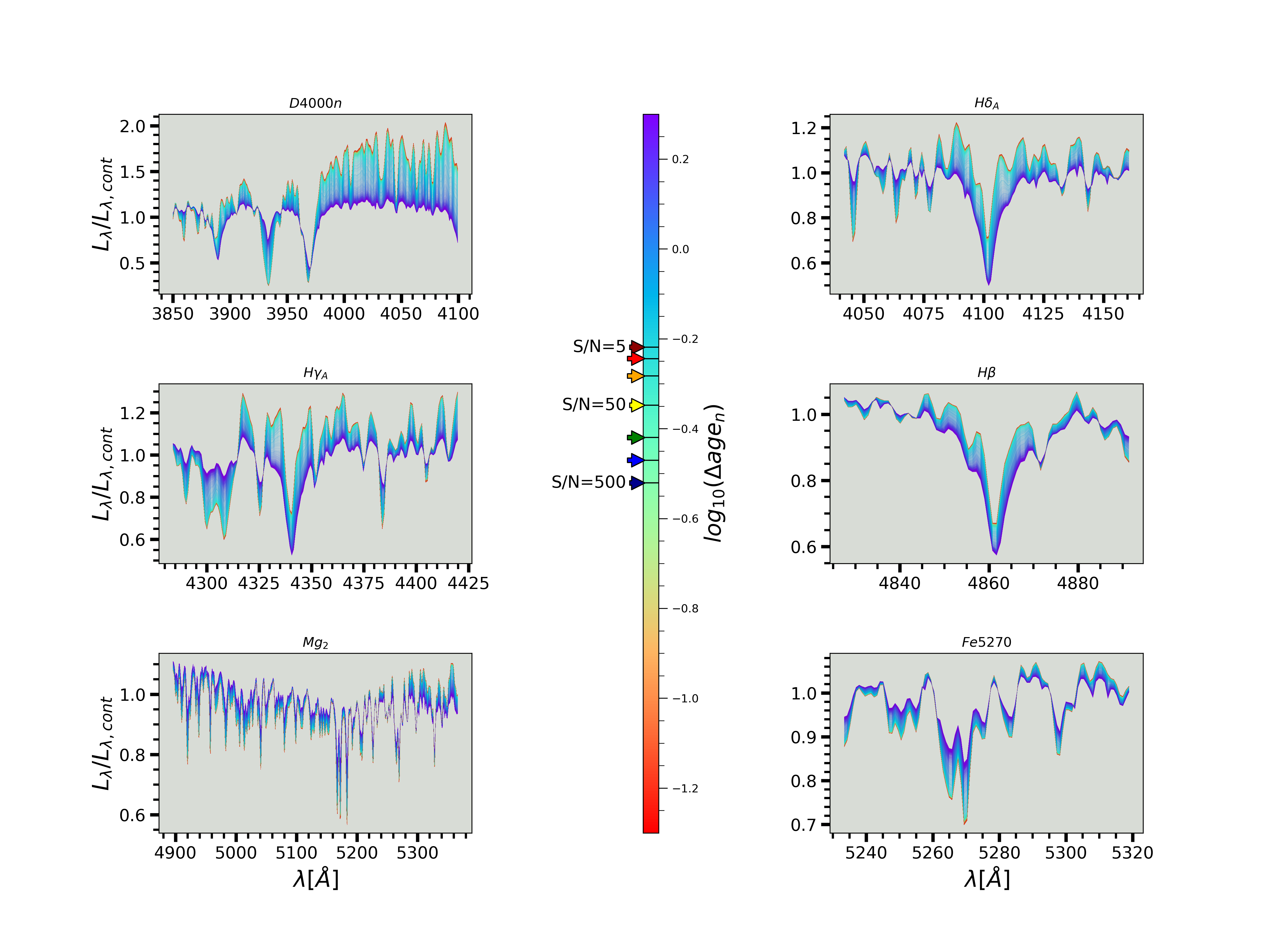}
    \caption{
    Spectral variations as a function of the relative SFH extension \dagen~in the wavelength regions corresponding to the most significant spectral indices, for the models characterized by $\log(\rm age_{50}/yr)$ between 9.5 and 9.55. The colour of each spectrum encodes its \dagen~value, according to the colour bar.
    The spectra are normalized to the mean flux in the two side-bands of each index, except for the $\rm D4000_n$ for which the normalization is done in the blue side-band.
    As in Fig. \ref{fig:stacked_specs}, the painted arrows next to the colour bar represent $\log($\dagenmin$)$ for different signal-to-noise ratios: Dark red: SNR=5\invAA; red: SNR=10\invAA; orange: SNR=20\invAA; yellow: SNR=50\invAA; green: SNR=100\invAA; blue: SNR=200\invAA; dark blue: SNR=500\invAA.}
    \label{fig:stacked_spec_indices}
\end{figure*}

In Fig. \ref{fig:stacked_spec_indices} we zoom in the spectral regions corresponding to six representative spectral absorption indices which are used as constraints (either alone or as part of composite indices), and plot the spectra colour-coded by \dagen~ as in Fig. \ref{fig:stacked_specs}. In order to ease the comparison, the spectra are normalized so to have the same average flux density in the index side-bands, except for the $\rm D4000_n$ index, for which the spectra are normalized to the blue side-band ($3850$\AA$-3950$\AA). The systematic behaviour of the absorption features with \dagen~is immediately apparent from these plots. We notice no appreciable variation in the absorption until \dagen$\approx 10^{-0.5}-10^{-0.3}$ is reached. For longer SFH duration, all three Balmer absorptions become more intense\footnote{This trend is reversed for ages $\lesssim10^{8.5}$~yr, where the time derivative of the Balmer indices reverses (see e.g. figure 19 in \citealt{BC03} and Fig. \ref{fig:index_time_deriv} below).} and the $\rm D4000_n$ decreases. Metal absorptions have in general a much more limited response to \dagen, with $\rm Fe5270$ and the H+K Ca absorptions around 3950\AA~becoming mildly weaker at larger \dagen~and $\rm Mg_2$ being essentially insensitive. From this figure we can appreciate that only a limited number of key spectral features over limited spectral ranges are sensitive to the duration of the SFH. This is part of the reason why the full-spectral fitting approach does not enhance the attainable age resolution in general, unless very high SNR is considered. 

\subsection{Linking age resolution and time evolution of SP spectral features}\label{sub:disc_spec_feat}
\begin{figure*}
 \centering
    \includegraphics[width=\textwidth]{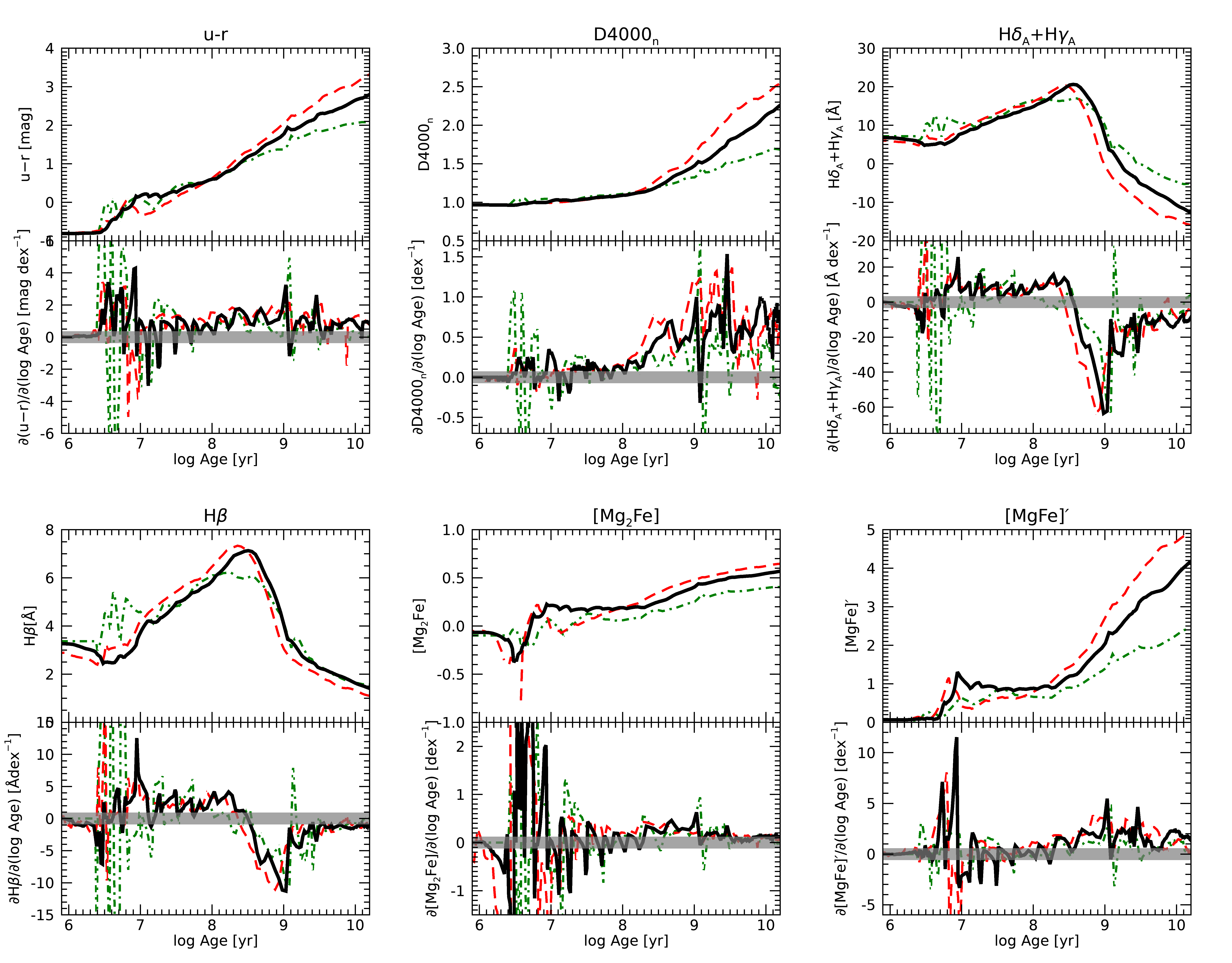}
    \caption{Age dependence (\emph{top panels}) and logarithmic age derivative as a function of age (\emph{bottom panels}) of the $u-r$ colour and of the five absorption indices used in our analysis, respectively. The black thick solid line is for $Z=Z_\odot$, the red long-dashed line for $Z=2.5\,Z_\odot$, and the green dot-dashed line for $Z=0.2\,Z_\odot$. The grey shaded line highlights the zero derivative. As discussed in the text, the ages of maximum amplitude in log derivative are typically associated with the best age resolution.}
    \label{fig:index_time_deriv}
\end{figure*}

From first principles, the limiting capability of resolving the duration of star-formation episodes must depend on how quickly the spectral features of a stellar population vary as a function of time. Therefore, in order to interpret the complex fine structure of the age resolution plots presented in the previous sections (Fig. \ref{fig:ageres_maps}, \ref{fig:ageres_snr_all} and \ref{fig:FSF_Zsolar}), we analyse how the five indices and the colours considered in this work vary as a function of age for the individual SSPs adopted as the base of our model library of composite stellar populations.
In the six panels of Fig. \ref{fig:index_time_deriv} we show the age dependency of each spectral feature (\emph{top graph}) and of its logarithmic age derivative (\emph{bottom graph}), for the $u-r$ colour and the five absorption indices, respectively. For better readability, only three metallicities are represented: solar ($Z/H \equiv 0.02$, solid black line), 0.2 times solar (dot-dashed green line) and 2.5 times solar (dashed red line). The logarithmic age derivative of the observed feature is indeed very tightly related to the relative \dagen~adopted to quantify the age resolution, in that, for a given spectral feature $x$, $\frac{\partial x}{\partial \log {\rm Age}} \propto\frac{\partial x}{\partial{\rm Age}/{\rm Age}} \simeq \frac{\Delta x}{\Delta{\rm Age_n}}$. Therefore we expect to find the best age resolution at the ages were the logarithmic age derivatives of the spectral features have the maximum amplitude. In fact, we can observe that all spectral features and, most strongly, the Balmer indices present a peak in the derivative at $\approx 10^{9.05}$~yr, which corresponds to the main peak in age resolution in any observing condition. Similarly, a peak in the derivatives of colours, $D4000_n$ and Balmer indices determines a peak in resolution at $\approx 10^7$~yr. At relatively low SNR ($\lesssim 20$\invAA) only these features seem to drive the variations in time resolution. When the SNR increases, the weaker and less age-sensitive metal features start to play a role and determine the conspicuous number of age-resolution peaks visible in the plots for SNR$\gtrsim 50$\invAA.

This connection between the fine structure of the age resolution function and the rapid variations of spectral features in the SSPs raises some caveats related to theoretical uncertainties and model-dependent results. While the secular trends of the indices as a function of time can be considered a robust prediction of any SP model, modulo small systematic offsets among different model renditions, the short time-scale fluctuations depend on the fine details of the model prescriptions and adopted stellar libraries. As a consequence, we recommend to take the fine structure of the age resolution function (especially for high SNR) with a grain of salt and refer to the smooth underlying trends as the robust result. The only exception to this general conclusion are the resolution peaks at $\simeq 10^9$ yr and $10^7$ yr, which are physically justified by well defined transitions in the evolutionary trends of most spectral features.

\subsection{How do age-resolution fluctuations affect SFH reconstruction?}\label{sub:disc_SFHreconstr}
\cite{Ocvirk+2006} thoroughly investigated the problem of spectral inversion into SFH from a mathematical point of view, based on the singular value decomposition (SVD) of the SSP spectral matrix. In this way, they were able to demonstrate the ill-posed nature of the problem and the link between the properties of the eigenvectors and the effective age resolution. In particular, they showed how the properties of the eigenvectors can bias the recovered SFH by, e.g., producing spurious bursts at particular ages (see their figure 7). Irrespective of the actual inversion method, it is quite common for the non-parametric reconstructed SFH to be quite bumpy rather than smooth and continuous. Examples of this behavior can be seen in figure 12 of \cite{cid_fernandes+13}, where the spatially resolved SFH derived with \textsc{starlight} for three different CALIFA galaxies all display peaks at $\simeq 10^9$ yr and $10^7$ yr. Similarly, \citet[figure 3]{Gonzalez-Delgado:2017aa}, based on \textsc{starlight} too, obtain clearly bimodal SFHs for all classes of galaxies, with one peak at $\simeq 10^9$~yr, and another one at the maximum possible age.  Multi-modal SFHs are also recovered by \cite{Sanchez:2019aa} for MaNGA galaxies based on \textsc{pipe3d} and by \cite{Cappellari:2023} for LEGA-C galaxies at $z\simeq 0.7$ with \textsc{pPXF}. 
The recurrence of these multiple modes at roughly the same characteristic ages for different galaxies across space and cosmic time suggests that they (or at least part of them) may be produced by method biases. Indeed, the work by \cite{boecker+20} lends further support to this hypothesis: by comparing the age-metallicity distribution of the stars in the nuclear star cluster in M54 obtained via resolved stellar population analysis and integrated-light analysis (with \textsc{pPXF}), one can see that the continuous distribution of ages between 1 and 7 Gyr found by the resolved analysis tends to be concentrated by the integrated analysis in a single peak around 1 Gyr. 

The recurrence of the SFH peak at $\simeq$1 Gyr, as well as the coincidence of other peaks (most notably at $10^7$ yr) with the maxima in the age derivative of the indices (hence in the age resolution) derived in this work, suggest that these particular ages may act as sort of ``attractors'' for the solution of the inversion problem. Weight appears to pile up at the transition ages at which the spectral features change more rapidly. In terms of SVD, recalling the analysis by \cite{Ocvirk+2006}, this effect can be interpreted as eigenvectors characterizing these transition ages to be always highly ranked. Although speculative, this hypothesis is worth further investigation in future works. If confirmed, this would indicate that non-parametric methods of SFH reconstruction may be prone to hidden biases, which are essentially hard-wired in the base SSPs, hence in the stellar physics. To take advantage of the inherent flexibility allowed by a non-parametric approach, it would then be relevant to explore ways to balance these possible biases.






\section{Conclusions}\label{sec:conclusions}

In this paper we have conducted a series of idealized experiments to quantify the age resolution (i.e. the minimum relative duration of a star-formation episode $\Delta {\rm age}_{n} \equiv \frac{\Delta {\rm age}_{10-90}}{{\rm age}_{50}}$) that is possible to measure via spectral analysis. By adopting simplified shapes for the SFH, fixed metallicity and no dust, we have shown that an age resolution of $0.4\lesssim$\dagenmin$\lesssim 0.7$ is achievable over most of the age-metallicity plane, except for ages less than a few 10-million years, where the resolution degrades substantially (or \dagenmin~becomes even impossible to measure). If this minimum resolution is adopted to define the bin size for reconstructing a SFH, this would translate in logarithmic bins between 0.15 and 0.25 dex wide. As a corollary of this conclusion, uncertainties in absolute age determinations cannot get significantly smaller that 0.1 dex either.
To this goal, good SNR spectroscopy (${\rm SNR}>20$\invAA) is required, with optimal results being reached at ${\rm SNR}\gtrsim 100$\invAA, beyond which the resolution saturates. We have shown that broad-band SED constraints or low-SNR ($<20$\invAA) spectroscopy is insufficient to resolve SFH durations shorter than 60\% of the median age for ages above 1 Gyr, and durations of the order of the median age for younger populations. On the other hand, using either spectral indices or the full spectrum does not make substantial difference. In particular, at low SNR indices are to be preferred to the full (rectified, i.e. continuum-normalized) spectrum, as it turns out that the additional spectral information carried by the full spectrum is more than counter-balanced by the additional noise. Only at very high SNR ($\gtrsim 100$\invAA) the full spectrum offers a real, although mild, advantage over the indices. We have shown that, at ${\rm SNR}\lesssim 20$\invAA, using the spectral shape in addition to the high-frequency absorption features may improve the resolution based on the full spectrum, especially at low ages and low metallicity, where a decrease of \dagenmin~by up to 0.15 dex can be observed. However, this result should be taken as an upper limit, as correlated noise was not properly treated in our analysis (and it is hard to treat for fitting codes in general), so that the constraining power of the spectral shape is dramatically over-estimated.

We have investigated the drivers and the physical origin of the varying age resolution at different ages and metallicities. As suggested by simple physical arguments, the age resolution is driven by the amplitude of the time derivative of the spectral features. Noteworthy, the peaks in age resolutions tend to appear as peaks in the reconstructed SFH in full-inversion non-parametric algorithms. This should be further investigated in future works as it might hint at possible hidden and hard-to-quantify biases in the reconstruction of complex SFHs. In turn, this should possibly encourage to reconsider advantages and disadvantages of parametric vs non-parametric approaches.

As a final consideration in the perspective of the archaeological reconstruction of SFHs, our results put a hard limit on their accuracy: no SF episode shorter than $\simeq 40\%$ of its median age can be resolved. This poses strong limitations to our ability to discern between smooth and bursty SFH and to estimate the duration of star formation or quiescence episodes that may mark phenomena such as interactions, mergers or AGN outbursts. Attempts at resolving the early (i.e. at look-back time $\approx 10$~Gyr) SFH of nearby galaxies are intrinsically limited to episodes longer than $\approx 4$~Gyr, which is largely insufficient to understand their evolution. Even in the simplest case of massive passive galaxies, the archaeological reconstruction of their SFH is essentially unable to provide any information about what happened at $z\gtrsim 1.5-1.7$, i.e. during the time lapse in which all the relevant action happened before quenching. This demonstrates the need to push the archaeological characterization of galaxies to high redshift by means of deep spectroscopy. Already at $z\simeq 1.5$ deep spectroscopy should enable a proper reconstruction of the SFH histories of galaxies across the cosmic noon, through the peak of activity and the quenching phase for the most massive ones. The advent of MOONS at the VLT \citep{MOONS} will open up this window of opportunity in the next years. In the meanwhile, deep surveys at intermediate redshift such as LEGA-C \citep[][$0.5\lesssim z \lesssim 1$]{LEGA-C}, and, at somewhat lower redshift ($z\sim 0.55$) the stellar population surveys StePS with WEAVE at the William Herschel Telescope \citep{WEAVE,Iovino+2023} and with 4MOST at the ESO-VISTA telescope \citep{4MOST,4MOST-StePS} will allow us to explore the range of cosmic time when galaxies experienced most of their star-formation quenching. Peering into the early evolutionary phases of galaxy formation will be enabled only once the James Webb Space Telescope will have gathered sufficiently large spectroscopic samples at high redshift, looking forward for the large spectroscopic surveys that will be possible only with MOSAIC at the ELT \citep{MOSAIC_sciencecase} in the next decades.

\section*{Acknowledgements}
We thank the anonymous referee for her/his constructive report, which has improved the clarity of our manuscript and motivated us to a more detailed exploration of several issues.

\section*{Data Availability}

Tables in electronic format are available for all results published in this paper (figures 2, 3, 4, and A1) on VizieR at the CDS \url{http://vizier.cds.unistra.fr}.



\bibliographystyle{mnras}
\bibliography{MN-23-2729-MJ.R1} 




\appendix

\section{Comparison of indices+colours vs full spectrum at low metallicity}\label{app:fullspec_20perc}
In this appendix, in order to complement the information provided in Sec. \ref{sec:results_full_spec}, we report the comparison of the limiting age resolution that can be obtained with indices+colours with respect to full spectral fitting, for subsolar metallicity ($Z=0.2\,Z_\odot$). The general trends are the same as at solar metallicity, but more marked differences between broad and narrow spectral ranges emerge at high SNR, while at low SNR $\simeq 5$\invAA~the advantage of using the spectral indices+colours instead of the full \emph{rectified} spectrum is amplified with respect to the solar metallicity case. 
\begin{figure*}
 \centering
    \includegraphics[width=\textwidth]{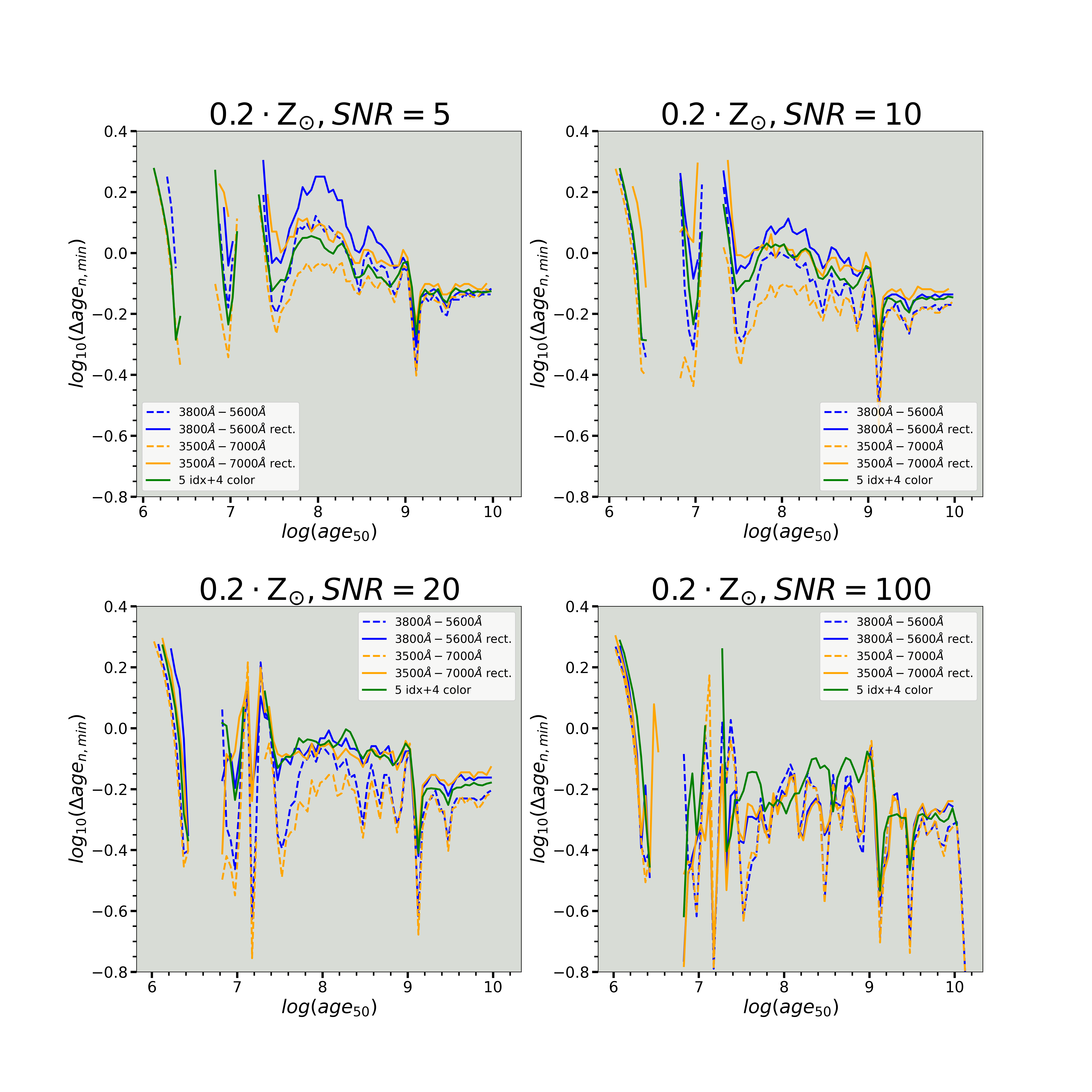}
    \caption{Comparison of the age resolution \dagenmin~achievable based on indices and colours (\emph{green line}) with that obtained from full spectral fitting, using two different wavelength ranges (narrow range in \emph{blue} and broad range in \emph{orange}), for $Z=0.2\,Z_\odot$ and four different SNR per \AA, as reported in each panel. Missing segments denote age bins in which the procedure for the computation of \dagenmin~failed, due to the chaotic evolution of spectral features around those ages.}
    \label{fig:FSF_20pecentZsolar}
\end{figure*}

\section{Recovering the relative duration of the SFHs in realistic cases}\label{app:dagen_recovery}
In this appendix we illustrate how well it is possible to constrain the relative duration of realistic SFHs, \dagen, using the Bayesian method of Zibetti et al. (\citeyear{Zibetti:2017aa}, Z17 hereafter, also \citealt{gallazzi+05}), which accounts for the degeneracies between SFH parameters, metallicity and dust attenuation, as obtained from the marginalized posterior PDF.

As in \citetalias{Zibetti:2017aa} we consider a comprehensive spectral library of 500\,000 models of complex SFHs. They are built considering a ``secular'' component for the SFH, that is represented by a \cite{Sandage:1986aa} function, on the top of which up to six instantaneous bursts are added, with different ages and intensities. Variable metallicity along the SFH is also implemented, following a simple ``leaking box'' function (see equation 3 in \citetalias{Zibetti:2017aa}) for the secular component and a randomized rendition of it for the bursts. Dust attenuation is implemented following the \cite{charlot_fall00} two-component prescription, with a diffuse ISM and a birth-cloud component, differently affecting stars of different ages. All parameters generating the models are randomly drawn to produce an extensive coverage of the space of physical parameters (e.g. mean age, SFH duration, metallicity and effective dust attenuation). More details are provided in \citetalias{Zibetti:2017aa}. For each model we compute the five indices introduced in Sec. \ref{sub:library} (for a velocity dispersion $\sigma=200$~km~s$^{-1}$) and the five SDSS broad-band fluxes, as well as the physical parameters, specifically the relative SFH duration \dagen.
A chunk of 12\,500 models is then taken to mock real observations: errors are attached to the indices assuming ${\rm SNR} =20$\invAA, while fixed errors to the broad-band photometric fluxes are attached, in the same way as described in Sec. \ref{sub:err_simul}. The remaining 487\,500 models are used to compute the marginalized posterior PDF of \dagen~for each of the mock galaxies, as explained in \citetalias{Zibetti:2017aa}.

\begin{figure*}
 \centering
    \includegraphics[width=\textwidth]{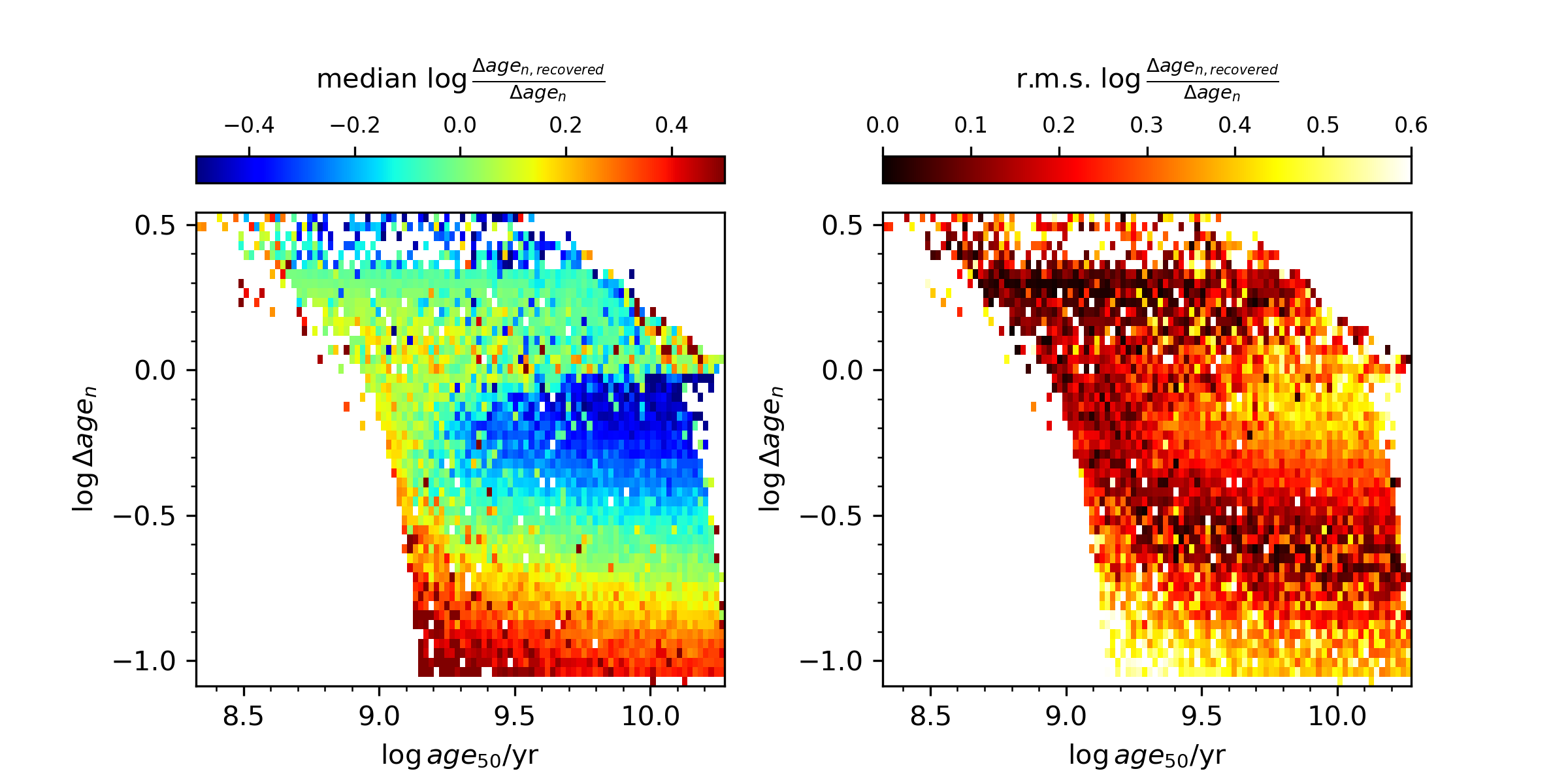}
    \caption{Distribution of the median bias (\emph{left panel}) and the r.m.s. (\emph{right panel}) of the recovered relative duration of the SFH, \dagen$_{\rm, recovered}$, with respect to the input true \dagen.}
    \label{fig:dagen_recovery}
\end{figure*}

For each mock galaxy, we consider as estimated \dagen$_{\rm recovered}$~the value corresponding to the median of the posterior PDF. We then compare these estimates with the true input value, by computing for each mock the logarithmic difference between the estimated and the true \dagen: $\log{\Delta{\rm age}_{\rm n, recovered}}-\log{\Delta{\rm age}_{\rm n}}=\log{\frac{\Delta{\rm age}_{\rm n, recovered}}{\Delta{\rm age}}}$. In Fig. \ref{fig:dagen_recovery} we plot the median (\emph{left panel}) and the r.m.s. (\emph{right panel}) of those differences in bins of true ($\rm age_{50}$, \dagen), for the 12\,500 mocks. As we can see, we can split the plane with a curved line going from ($\rm age_{50}\sim 1$\,Gyr, $\log$\dagen$\sim -0.3$) to ($\rm age_{50}\gtrsim 14$\,Gyr, $\log$\dagen$\sim 0$). In the upper part of the plane the estimated \dagen~are substantially unbiased with a scatter around 0.1-0.2 dex with respect to the true input value. Below the divide, the recovered \dagen~basically reflect the median of the model library prior, thus we see a transition from negative to positive bias moving down in the plane.

This test indicates that, considering realistic/plausible SFHs, chemical enrichment histories and dust, the capability of recovering the relative duration of a SFH is much more limited than the theoretical limit computed in this article: for most of the old stellar populations reliable \dagen~estimates are possible only as long as the duration of the SFH is comparable with the age itself. Moving down to ages around 1 Gyr, durations as short as roughly 1/2 of the age can be reliably estimated. Shorter durations are in general ill-determined.

\bsp	
\label{lastpage}
\end{document}